\documentstyle[psfig,12pt]{article}
\def\href#1#2{#2}

\def\beq{\begin{equation}}
\def\eeq{\end{equation}}

\begin{document}
\baselineskip=14.5pt
\pagestyle{plain}
\setcounter{page}{1}
\renewcommand{\thefootnote}{\fnsymbol{footnote}}
\newcommand{\da}{\dot{a}}
\newcommand{\db}{\dot{b}}
\newcommand{\dn}{\dot{n}}
\newcommand{\dda}{\ddot{a}}
\newcommand{\ddb}{\ddot{b}}
\newcommand{\ddn}{\ddot{n}}
\newcommand{\pa}{a^{\prime}}
\newcommand{\pb}{b^{\prime}}
\newcommand{\pn}{n^{\prime}}
\newcommand{\ppa}{a^{\prime \prime}}
\newcommand{\ppb}{b^{\prime \prime}}
\newcommand{\ppn}{n^{\prime \prime}}
\newcommand{\fda}{\frac{\da}{a}}
\newcommand{\fdb}{\frac{\db}{b}}
\newcommand{\fdn}{\frac{\dn}{n}}
\newcommand{\fdda}{\frac{\dda}{a}}
\newcommand{\fddb}{\frac{\ddb}{b}}
\newcommand{\fddn}{\frac{\ddn}{n}}
\newcommand{\fpa}{\frac{\pa}{a}}
\newcommand{\fpb}{\frac{\pb}{b}}
\newcommand{\fpn}{\frac{\pn}{n}}
\newcommand{\fppa}{\frac{\ppa}{a}}
\newcommand{\fppb}{\frac{\ppb}{b}}
\newcommand{\fppn}{\frac{\ppn}{n}}
\newcommand{\A}{A}
\newcommand{\B}{B}
\newcommand{\mmu}{\mu}
\newcommand{\mnu}{\nu}
\newcommand{\ii}{i}
\newcommand{\jj}{j}
\newcommand{\jl}{[}
\newcommand{\jr}{]}
\newcommand{\ml}{\sharp}
\newcommand{\mr}{\sharp}


\begin{flushright}
UW/PT 04-24\\
{\tt hep-th/0412017}
\end{flushright}

\vskip 2cm

\begin{center}
{\Large \bf Hawking and Page on the brane} \\
\vskip 1cm

{\bf A.~Chamblin$^1$ and A.~Karch$^2$} \\
\vskip 0.5cm
{\it $^1$  Department of Physics, University of Louisville,
Louisville, KY 40292}\\
{\tt E-mail: chamblin@mit.edu} \\
\medskip
{\it $^2$ Department of Physics, University of Washington,\\
Seattle, WA 98195-1560} \\
{\tt E-mail: karch@phys.washington.edu}  \\
\end{center}

\vskip1cm

\begin{center}
{\bf Abstract}
\end{center}
\medskip
We show that the Hawking-Page phase transition of a CFT on AdS$_{d-1}$ weakly
coupled to gravity has a dual bulk description in terms of a phase
transition between a black string and a thermal gas on AdS$_{d}$.
At even lower temperatures the black string develops a Gregory
Laflamme instability, which is dual to black hole evaporation in the
boundary theory.

\newpage

\section{Introduction}

Gravitational theories on Anti de-Sitter space undergo an
interesting phase transition as the temperature is varied, the
well known Hawking-Page phase transition \cite{hp}. At high
temperatures a large black hole is formed, whereas at low
temperatures a thermal gas in the AdS$_d$ space with curvature
radius $L$ is the preferred configuration. A first order
transition occurs between the two geometries at a temperature
$T_{HP}=\frac{d-2}{2 \pi L}$. Using AdS/CFT the corresponding phase
transition gets mapped to a confinement/deconfinement transition
in the CFT on the boundary. As typical in a system with a first
order transition a second interesting temperature $T_S=
\frac{\sqrt{(d-2)^2-1}}{2 \pi L}<T_{HP}$ is associated with the
``spinodal'', the point at which the false minimum disappears: In
the regime $T_S<T<T_{HP}$ both configurations are allowed, but the
thermal AdS is preferred. The black hole however still has a
positive specific heat, and like with water vapor lowered below
the boiling point, one can go to a supercooled quark gluon plasma
that is still stable against small thermal fluctuations and needs
to nucleate bubbles in order to transition to the global free
energy minimum. Below $T_S$ the specific heat of the black hole
becomes negative and the black hole configuration ceases to exist
even as a local minimum of the free energy.

Most of the discussion of the Hawking-Page phase transition in the
context of AdS/CFT focuses on the transition in the $d$
dimensional bulk and its $d-1$ dimensional field theory
interpretation. But AdS/CFT also allows us to study a conformal
field theory on AdS$_{d-1}$ as the {\it boundary geometry} \cite{kr}, weakly
coupled to gravity by introducing a UV cutoff in the form of an RS
Planck brane \cite{RS2}. In this setup the theory on the
boundary will undergo a Hawking-Page phase transition, and now one
can ask what the bulk description of both the phase transition and
the onset of the black hole instability at the spinodal point are
in terms of the bulk. In this paper we will show that former maps
to a novel Hawking-Page like transition in the bulk between a
black string and thermal AdS, while latter maps to classic
instability in the bulk of the Gregory Laflamme type \cite{GL}. A
similar treatment has appeared in \cite{emparan1,emparan2} and we
will discuss the relation between their findings and our results.

The main motivation is to test the gauge/gravity duality in the
context of cutoff AdS spaces like they appear in the RS scenario.
While AdS/CFT is well established in the conformal case, the
duality between a cutoff CFT coupled to gravity and the bulk
gravity-only theory has so far mostly been tested in the regime
where the localized graviton can be treated linearly. By studying
the Hawking-Page transition in this setup we can put this duality,
which is often invoked for model building purposes, on firmer
footing.

\section{The Hawking-Page Phase transition on the boundary}

\subsection{General Setup}

In order to establish the phase transition, one calculates the
difference of the on-shell gravitational action, \beq
\label{action} I = - M_{Pl}^{d-2} \int \sqrt{-g} d^dx (R -
2\Lambda), \eeq evaluated on the two solutions, which gets
identified with the difference in $\beta F$, where $F$ is the free
energy and $\beta$ the inverse temperature. The two configurations
are thought of as two saddle points that dominate the euclidean
path integral. Note that the contribution of matter fields to the
action is ignored. This is reasonable as long as $M_{Pl}$ is
larger than the other scales, so that the gravitational
contribution to the action dominates. Note in particular that as
dynamical gravity is switched off by taking Newton's constant to
zero, that is $M_{pl}$ goes to infinity, the difference in free
energies becomes extremely large and the path integral over
metrics will still be dominated by one or the other configuration,
so the Hawking Page transition persists in that limit. Of course
the time it takes to actually transition from one geometry to the
other will go to infinity as dynamical gravity is switched off.
This is precisely the situation that one encounters when choosing
to study AdS$_{d}$/CFT$_{d-1}$ with an AdS$_{d-1}$ metric on the
boundary. Writing AdS$_d$ in AdS$_{d-1}$ slicing where the metric
$ds^2_{AdS_{d-1}}$ has curvature radius $L$ as well, \beq
\label{ads} ds^2 = \cosh^2 (\frac{r}{L}) \, ds^2_{AdS_{d-1}} +
dr^2 \eeq one obtains as a dual a CFT on AdS$_{d-1}$, or more
precisely two copies of AdS$_{d-1}$ at $r=\pm \infty$, which
communicate with each other via their common boundary \cite{kr} as
we will review in detail in the next subsection.

Introducing a UV brane at $r=r_0$ serves as a UV cutoff in the
field theory and in addition introduces dynamical gravity with an
almost massless graviton, where the mass can be understood as
being a CFT loop effect \cite{porrati}. The curvature radius on
the brane is $l=\cosh(\frac{r_0}{L}) L$. So at least for $r_0 >>
L$ we are precisely in the regime where the Hawking-Page analysis
applies. Even when we completely remove the UV brane and the $d-1$
dimensional Planck scale goes to infinity so that the path
integral over metrics is just equal to a sum over saddle points,
we still have a different configuration dominating at low and high
temperatures. To actually make a transition from one to the other
would however use an infinite amount of time. In this paper we
will precisely consider this $r_0>>L$ regime of a boundary CFT on
AdS$_{d-1}$ weakly coupled to gravity. As we discussed, going
beyond this regime will result in significant changes to the
Hawking Page picture due to the matter effects. While a detailed
discussion of those is clearly beyond the scope of the present
paper, we can identify how they manifest themselves on the bulk
side.

\subsection{Boundary Conditions: Interplay of two CFTs}

The bulk metric only defines the conformal structure at the
boundary. By picking a defining function suggested by the slicing,
in our case $\cosh(\frac{r}{L})$, one picks a particular metric on
the boundary. This way different coordinate systems in AdS give
different metrics on the boundary, but they all have the same
conformal structure. The difference becomes physically once
conformal invariance is broken, e.g. in our case by the cutoff
brane(s) and by the finite temperature.

To understand better the case of interest in this paper, where the
boundary consists of two copies of AdS$_{d-1}$, one can first of
all consider the conformal case of pure AdS$_d$ in the bulk, where
the physics is equivalent to the standard CFT on the $S^{d-2}
\times R$ boundary. Many of the issues discussed here can be found
also in \cite{kr,porrati}. The two copies of AdS$_{d-1}$ can be
conformally mapped to the southern and northern hemisphere of the
$S^{d-2} \times R$ boundary respectively. The conformal boundary
of AdS$_d$ gets realized as two disjoint AdS$_{d-1}$ spaces which
communicate via boundary conditions, so that together they have
the right topology of the $S^{d-2} \times R$ boundary to which
they are conformally related. The boundary conditions that need to
be imposed on the two copies of AdS$_{d-1}$ follow directly from
this conformal map: on the $S^{d-2}$ sphere the equator is not a
special locus at all, all fields have to be continuous and
differentiable across it. When mapped to two copies of AdS$_{d-1}$
this implies that the value of the field and its derivative close
to the boundary of one AdS$_{d-1}$ determine the same information
for the same field at the same point of the $S^{d-3} \times R$
boundary of the other AdS$_{d-1}$. For a lightray hitting the
boundary one imposes transparent boundary conditions instead of
the usual total reflecting ones, that is the signal leaves one
AdS$_{d-1}$ and enters the other AdS$_{d-1}$. The holographic dual
in this case is either a CFT living on $S^{d-2} \times R$ or
equivalently (simply related by a conformal transformation): two
copies of AdS$_{d-1}$, each with its own CFT, communicating with
each other via the transparent boundary conditions we just
described.

As soon as one introduces the cutoff conformal invariance is
broken, the boundary theory becomes sensitive to the actual
boundary metric and not just the conformal structure. In the bulk
the difference between distinct coordinate systems becomes
physical -- a brane at $r=const.$ in AdS$_{d-1}$ slicing is at a
different location than a brane in the standard Minkowski slicing.
To discuss the most general scenario let us introduce two cutoff
branes, one at a positive value of $r$, $r=r_0$, and one at a
negative value of $r$, $r=-r_1$. The holographic dual in this case
is hence again two copies of AdS$_{d-1}$, each with its own CFT
and dynamical graviton. $r_0$ and $r_1$ set the curvature radius
of the two AdS$_{d-1}$ spaces respectively. The two theories
communicate with each other via the transparent boundary
conditions for all fields including the graviton. Each AdS$_{d-1}$
is hence described by $d-1$ dimensional gravity coupled to a $d-1$
dimensional CFT and is expected to undergo its own Hawking Page
phase transition. The only departure from the standard analysis is
that instead of the reflecting boundary conditions here we have
the transparent boundary conditions that link the two AdS$_{d-1}$
spaces. Naively one would expect that this does not influence the
properties of the black hole: as long as the temperature in the
two AdS$_{d-1}$ spaces is the same, it should not matter if the
black hole is in a thermal bath of gravitons that get reflected
from the boundary or in a thermal bath of gravitons at the same
temperature that freely move from one AdS$_{d-1}$ to the other.
This expectation will be born out by computations in this paper.

One interesting special case is $r_1 \rightarrow \infty$, in which
case dynamical gravity switches off in one of the two AdS$_{d-1}$.
Another one is $r_1=r_0$. In this case one can impose an orbifold
projection $r \rightarrow -r$. In the orbifolded theory the
boundary is just one copy of AdS$_{d-1}$ and the transparent
boundary conditions turn into the regular reflecting boundary
conditions. Both those special cases are completely consistent
with the analysis in this paper.

Last but not least let us emphasize that even in the limit that
both cutoffs are taken to infinity one does not simply recover the
standard bulk Hawking Page physics. In the standard scenario one
demands the metric to take the form $S^{d-2}$ $\times$ $S^1$ close
to the boundary, corresponding to a CFT on the sphere at finite
temperature. Since we are interested in the scenario where gravity
on the AdS$_{d-1}$ slices is dynamical, we allow for non-trivial
geometry on those slices and only fix the boundary conditions on
the $S^{d-3}$ $\times$ $S^1$ along which the two AdS$_{d-1}$
components of the boundary communicate. While in the cutoff to
infinity limit dynamical gravity decouples, one of course only
obtains a smooth limit if one does not suddenly change the
boundary conditions. In particular the black string solution we
will introduce momentarily, which gives rise to two boundary
AdS$_{d-1}$ Schwarzschild black holes, would be allowed by our
boundary conditions and not by the standard bulk Hawking Page
analysis. As is apparent from formula (\ref{free}) the free energy
of the black string is negative and grows exponentially in the
cutoff $r_0$, whereas the free energy of the large AdS$_d$ black
hole is independent of the cutoff.  As soon as the boundary
conditions allow for the black string, it completely dominates the
thermal ensemble for large values of the cutoff and the physics is
genuinely different from the standard bulk Hawking Page analysis.

\subsection{The high temperature phase -- The black string solution}

The Hawking Page transition on the boundary occurs when the
inverse boundary temperature is $\beta^{bound}=\frac{2 \pi
l}{d-3}$. At higher temperatures the dominant configuration should
yield a $d-1$ dimensional AdS Schwarzschild (S-AdS) black hole as
the boundary metric. Or to be precise we get again two copies of
this spacetime at $r=\pm \infty$; for the reasons discussed in the
last subsection, we will focus on the physics associated with one
of those black holes.

One solution that realizes this boundary geometry is the black
string \cite{chamblin}
\begin{equation}
\label{sads}
ds^2 = \cosh^2 (\frac{r}{L}) ds^2_{S-AdS_{d-1}} + dr^2.
\end{equation}
That is one simply replaces the AdS$_{d-1}$ with a S-AdS$_{d-1}$
metric on every slice, as depicted in Figure 1. While for flat or
dS slices this string suffers from a Gregory Laflamme instability,
it has been argued in \cite{emparan1} based on thermodynamic
arguments that for sufficiently large horizon radius $r_H$ the
S-AdS black string is actually stable. This has been confirmed by
a numerical analysis for $d=5$ in \cite{numerics}. $T_{HP}$
corresponds to $r_H=L$. In order to establish that the black
string is the dominant configuration for $r_H>L$, one can use the
euclidean path integral as in the original Hawking Page paper to
establish that the difference in free energies associated with the
metrics (\ref{ads}) and (\ref{sads}) is given by the difference in
the value of on-shell actions (\ref{action}) \beq \label{free}
\beta  \Delta{F} = I^{Black-String} - I^{AdS_d} =
\int_{-r_1}^{r_0}dr \, \cosh^{d-3} (r/L) \left ( I^{S-AdS_{d-1}} -
I^{AdS_{d-1}} \right ). \eeq But $I^{S-AdS_{d-1}} - I^{AdS_{d-1}}$
is just the free energy difference that is governing the $d-1$
dimensional Hawking Page transition on the brane. It is negative
for $r_H>L$ and positive for $r_H<L$. To the extend that these are
the only two geometries that correspond to a boundary CFT on AdS
at a given temperature, this establishes that in the high
temperature regime the black string is the dominant configuration.
We will address the issue of possible other geometries later.

Note that the sign of this free energy difference is completely
independent of $r_1$. Both the black hole radius on the slice as
well as the AdS$_{d-1}$ curvature radius of the slice grow with
the same power of the warpfactor. The difference in free energies
is either positive on all slices or negative on all slices. For
generic values of $r_0$ and $r_1$ the phase transition occurs when
on every slice, including the two cutoff branes, the black hole
radius is equal to the AdS$_{d-1}$ curvature radius of the slice.
In the special case of $r_0=r_1$ with the orbifold projection
imposed, corresponding to the standard reflecting  boundary
conditions in AdS$_{d-1}$ as we explained above, we can take the
lower integration boundary to be $r=0$. This confirms our
expectation that the transparent boundary conditions do not alter
the $d-1$ dimensional Hawking Page physics at all.

\begin{figure}
   \centerline{\psfig{figure=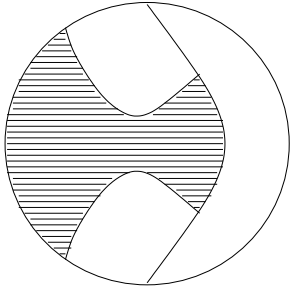,width=1.5in}}
    \caption{The black string solution.}
\label{string}
 \end{figure}

\subsection{The low temperature phase -- Thermal AdS}

At temperatures below $T_{HP}$ the dominant boundary metric should be thermal AdS$_{d-1}$, that
is AdS$_{d-1}$ with periodic euclidean time. This is what
we get by simply replacing the bulk AdS$_d$ with thermal AdS$_d$. To compare temperatures,
we have to examine the change of variables between (\ref{ads}) and a more conventional way
of writing euclidean AdS$_d$, e.g.
\beq
\label{global}
ds^2 = \cosh^2(\frac{\rho}{L}) dt^2 + d \rho^2 + \sinh^2(\frac{\rho}{L}) d \Omega_{d-2}^2.
\eeq
The change of variables has for example been presented in
\cite{kr}. The important point is that $t=\tilde{t}$ and hence
$\cosh(\frac{r}{L}) \cosh(\frac{\tilde{\rho}}{L}) =
\cosh(\frac{\rho}{L})$. Tilde variables refer to coordinates of
the $d-1$ dimensional AdS on the boundary when written in a global
coordinate system analogous to (\ref{global}). In the context of
AdS/CFT the relevant notion of boundary temperature is that if the
metric in the bulk has the asymptotic form $f^2 ds^2_{bound.}$
where $ds^2_{bound.}$ is the boundary metric and $f$ a function of
the coordinates that diverges linearly as we approach the
boundary, the inverse temperature is the period of the time
variable with asymptotic metric $f^2 dt^2$. In our case the
temperature of the boundary CFT has to be measured at the $S^1
\times S^{d-3}$ that separates the two AdS$_{d-1}$ components of
the boundary of AdS$_d$. In the coordinates (\ref{global}) it is
time times the equator of the $S^{d-2}$.

We see that on the brane with curvature radius
$l=\cosh(\frac{r_0}{L}) L$ a period $\beta^{bulk}$ of the $t$
coordinate implies an inverse temperature \beq \beta^{bound} =
\cosh(\frac{r_0}{L}) \beta^{bulk} \eeq and correspondingly for
$r_1$. So at the phase transition point, the thermal boundary
AdS$_{d-1}$ with $\beta^{bound} =\frac{2 \pi l}{d-3}$ is what we
obtain from a thermal bulk AdS$_d$ with \beq \label{bulk}
\beta^{bulk} = \frac{2 \pi L}{d-3}. \eeq The Hawking Page phase
transition on the boundary maps to a very similar phase transition
in the bulk, where at high temperatures a black string dominates
the euclidean path integral, whereas at low temperatures thermal
AdS is the relevant configuration.

\subsection{The bulk black hole does not contribute}

In order to argue that this is the complete picture for the ensemble we study, we have to make
sure that there are no other configurations that yield the same $S^1 \times S^3$ geometry on
the boundary of the boundary and have an even lower free energy.
One obvious candidate is the bulk black hole, S-AdS$_d$. First note that the value
of $\beta$ at which the usual {\it bulk} Hawking Page transition happens, $\beta^{HP} = \frac{2 \pi L}{d-2}$,
is lower than the value relevant for the boundary Hawking page transition (\ref{bulk}),
and hence the corresponding temperature is higher.
So by the time the boundary made the transition to thermal AdS$_{d-1}$, in the bulk the S-AdS$_d$ black hole
is already a subdominant configuration -- actually the temperature is even
lower than $T_S$, so the bulk black hole solution doesn't even exist anymore.

S-AdS$_d$ could still be relevant for the high temperature regime.
For the case $r_0, r_1>>L$ we are discussing it is however clear
that the black string will always dominate over the black hole.
The difference in free energies between the string and the
background AdS according to (\ref{free}) grows exponentially with
$r_0$ and $r_1$, while the difference between the bulk S-AdS$_d$
and the background approaches a fixed value independent of $r_0$
and $r_1$. The same argument also suppresses other geometries that
would lead to induced metrics on the brane that differ from
thermal AdS and the S-AdS$_{d-1}$ black hole, like the bulk black
hole which gives rise to an FRW like cosmology on the brane. For
$r_0, r_1>>L$ only the same two geometries as in the original
Hawking-Page transition contribute on the brane, and the two bulk
geometries we considered are the bulk extensions corresponding to
those two boundary geometries. If we want to go beyond the $r_0,
r_1 >>L$ regime, the bulk black hole certainly will start
contributing. On the boundary side this maps to the strong
correction due to matter effects becoming important.

\section{Quantum Corrections}

The induced metric in the high temperature phase is exactly that
of a lower dimensional AdS Schwarzschild black hole. In
\cite{emparan2} however it was shown that a brane world black hole
can {\it not} be a solution to vacuum Einstein equations on the
brane. AdS/CFT predicts that classical gravity in the bulk is dual
to the quantum CFT including planar corrections on the boundary,
and coupled to classical gravity if the UV brane is
introduced. In particular $$M_{Pl,d-1}^{d-3} = (M_{Pl,d} L)
M_{Pl,d}^{d-3},$$ where $M_{Pl,d} L$ is a large number, e.g.
$N^{2/3}$ in Maldacena's $d=5$ example. So for
$M_{Pl,d}<<M<<M_{Pl,d-1}$ bulk Hawking radiation is negligible,
while on the brane Hawking radiation and its backreaction
dominate\footnote{The authors of \cite{emparan2} were mostly
interested in $d=5$ black holes with $M> (M_5 L)^{\frac{3}{2}} M_4$,
which are those with a horizon larger than $L$ and have a right to
be called brane black holes and Hawking radiation only yields a
small correction in both 5 and 4 dimensional description. But the
argument that bulk Hawking radiation can be neglected did not rely
on this.}

{} From this point of view it is a puzzle why the black string
gives a metric on the brane that is not corrected at all due to
Hawking radiation. While this still remains a puzzle to us in
general, in the $r_0, r_1>>L$ regime we are analyzing, where the
UV brane is close to the boundary this is not an issue. The 4d
Planck scale is still $M_{Pl,4}^2=M_{Pl,5}^3 L$, whereas the 4d
curvature radius on the brane is $l = \cosh( \frac{r_0}{L})$, so
for $r_0>>L$ a black hole with horizon radius larger than $l$ is
exponentially heavier than the Planck scale and any backreaction
from Hawking radiation can be neglected. The same reasoning also
applies to $r_1$. In the general case the black string solution
suggests that the large S-AdS blackhole solution has all effects
of Hawking radiation included. This is also the picture one is led
to from the original Hawking-Page analysis, which suggests that
the large black hole is in thermal equilibrium with the heat bath
generated by its own Hawking radiation.

\section{The onset of the Gregory Laflamme instability}

Like the bulk, the boundary black hole also disappears at a
spinodal at temperature $T_S^{bound}$. While in the temperature range
$T_s^{bound}<T<T_{HP}^{bound}$ the thermal AdS is the dominant configuration
both in the bulk and on the boundary, the black hole solution on
the boundary is still a valid local minimum of the free energy.
But when the temperature is lowered below $T_S^{bound}$ the black hole
on the boundary has now a negative heat capacity and no longer
even represents a local minimum. At the same time the Gubser-Mitra
hypothesis \cite{GM}, which states that classical stability of a
black string should come hand in hand with local thermal
stability, would now say that the black string in the bulk should
develop a classical instability of the Gregory Laflamme type.
While the numerical analysis of \cite{numerics} is too crude to
give a decisive answer to this question about the precise value at
which the instability sets in, their result is consistent with
this interpretation.

On the boundary, the black hole will start
evaporating, and will eventually disappear (as long as we keep the
induced Planck scale large but finite). At least the initial stage
of this evaporation process is completely captured by the
classical evolution of the unstable mode in the black string
background. Once the black string becomes sufficiently hot, $d$
dimensional quantum effects will become important and classical
bulk gravity is no longer a good description, so the final stages
of the evaporation process go beyond classical gravity.

This is an explicit realization of the scenario first envisioned
in \cite{chamblin}, where the classical dynamics associated with
the black string in the bulk captures the physics of the black
hole on the brane. Additional supporting evidence for our picture
comes from the analysis in $d=4$, where the so called c-metric
yields an exact solution describing the snapping string
\cite{emparan1}. Based on thermodynamic arguments they find that
for small enough mass (which for the negative specific heat black
holes means large temperature) the black string becomes
disfavored. Instead there exist two classically stable black hole
configurations on the brane. These should be thought of as the
endproducts of the classical evolution of the black string, any
further evaporation process would be due to bulk quantum effects.

One can envision various scenarios for how precisely the black
string evolves\footnote{As pointed out to us by N. Kaloper and R.
Emparan, a snapping string does not seem to describe Hawking
radiation escaping to infinity -- either in the field theory the
radiation stays close to the black hole during the initial stages
of the evaporation until bulk quantum effects set in or the string
in addition to shrinking at its waist will slide of the brane
towards the IR in order to describe the expanding cloud of Hawking
radiation.}. One still somewhat controversial aspect is whether
the string snaps at all \cite{hm}. If indeed the final
configuration of the string is a stable non-uniform string, we
would get a stable remnant as the endresult of the evaporation on
the brane. In any case, the map to quantum black hole physics
really shows how important it is to understand what happens at the
Gregory Laflamme point. It is however safe to say that whatever
the string does --- that is precisely what the black hole is going
to do on the brane!

\section*{Acknowledgements}

We thank Roberto Emparan, Nemanja Kaloper and Toby Wiseman for
very helpful discussions about black holes on the brane and for comments on
the draft. The research of AK is supported in part by DOE contract
\#DE-FG03-96-ER40956 and AK is also supported by NSF Grant SBE-0123552
as an ADVANCE professor.

\bibliography{hp}
\bibliographystyle{ssg}
\end{document}